\documentclass[conference]{IEEEtran}
\IEEEoverridecommandlockouts
\usepackage{cite}
\usepackage{amsmath,amssymb,amsfonts}
\usepackage{algorithm}
\usepackage{algpseudocode}
\usepackage{graphicx}
\usepackage{textcomp}
\usepackage{xcolor,colortbl}
\usepackage{enumitem}
\usepackage{array}
\usepackage{soul}
\usepackage{subcaption}
\usepackage{lipsum}
\usepackage[font=small]{caption}
\newcommand{\PreserveBackslash}[1]{\let\temp=\\#1\let\\=\temp}
\newcolumntype{C}[1]{>{\PreserveBackslash\centering}p{#1}}
\newcolumntype{R}[1]{>{\PreserveBackslash\raggedleft}p{#1}}
\newcolumntype{L}[1]{>{\PreserveBackslash\raggedright}p{#1}}
\usepackage[none]{hyphenat}
\usepackage{blindtext}
\usepackage{multirow}
\usepackage{booktabs}
\usepackage{afterpage}
\usepackage[hidelinks]{hyperref}

\AtBeginDocument{%
    \setlength\intextsep{3pt}%
    \setlength\textfloatsep{3pt}}

\DeclareCaptionFont{Small}{\fontsize{8}{8}\selectfont}
\def\BibTeX{{\rm B\kern-.05em{\sc i\kern-.025em b}\kern-.08em
    T\kern-.1667em\lower.7ex\hbox{E}\kern-.125emX}}
\begin{document}
\begin{titlepage}\centering
\vspace*{\fill}
This work will appear in the proceedings of the 2024 IEEE Conference on Field Programmable Logic and Applications.
\vspace*{\fill}
\end{titlepage}

\title{
H2PIPE: High Throughput CNN Inference on FPGAs with High-Bandwidth Memory
}

\author{
\IEEEauthorblockN{Mario Doumet\IEEEauthorrefmark{1}, Marius Stan\IEEEauthorrefmark{2}, Mathew Hall\IEEEauthorrefmark{2}, Vaughn Betz\IEEEauthorrefmark{1}\IEEEauthorrefmark{3}}
\IEEEauthorblockA{\IEEEauthorrefmark{1} Department of Electrical and Computer Engineering, University of Toronto, Toronto, Canada \\
\IEEEauthorrefmark{2} Microsoft, Redmond, USA  \IEEEauthorrefmark{3}Vector Institute, Toronto, Canada}
E-mails: mario.doumet@mail.utoronto.ca, \{mariusstan, mathewhall\}@microsoft.com, vaughn@ece.utoronto.ca
}
\maketitle

\begin{abstract}
Convolutional Neural Networks (CNNs) combine large amounts of parallelizable computation with frequent memory access. Field Programmable Gate Arrays (FPGAs) can achieve low latency and high throughput CNN inference by implementing dataflow accelerators that pipeline layer-specific hardware to implement an entire network. By implementing a different processing element for each CNN layer, these layer-pipelined accelerators can achieve high compute density, but having all layers processing in parallel requires high memory bandwidth. Traditionally this has been satisfied by storing all weights on chip, but this is infeasible for the largest CNNs, which are often those most in need of acceleration. In this work we augment a state-of-the-art dataflow accelerator (HPIPE) to leverage both High-Bandwidth Memory (HBM) and on-chip storage, enabling high performance layer-pipelined dataflow acceleration of large CNNs. Based on profiling results of HBM's latency and throughput against expected address patterns, we develop an algorithm to choose which weight buffers should be moved off chip and how deep the on-chip FIFOs to HBM should be to minimize compute unit stalling. We integrate the new hardware generation within the HPIPE domain-specific CNN compiler and demonstrate good bandwidth efficiency against theoretical limits. Compared to the best prior work we obtain speed-ups of at least 19.4x, 5.1x and 10.5x on ResNet-18, ResNet-50 and VGG-16 respectively. 
\end{abstract}   

\fontsize{9.8}{11}\selectfont

\section{Introduction}
Over the past decade, the computational demands of Convolutional Neural Networks (CNNs) have increased rapidly, often requiring specialized hardware to meet performance requirements. FPGAs have emerged as a high-performance solution and AI-Optimized FPGAs have demonstrated capabilities even beyond those of a GPU on certain AI workloads \cite{beyond}. There are two main computational paradigms to accelerate CNNs on FPGAs \cite{survey}. The first is a \textit{Processing Element} (PE) based architecture which relies on one or more general-purpose convolution units that are sufficiently flexible to process any layer; these accelerators typically process one layer at a time \cite{dla}. The second type is a \textit{dataflow} style architecture, which assigns layer-specialized processing elements to all layers of a CNN, and computes multiple (often all) layers in a pipelined parallel fashion~\cite{xilinx_dataflow, hpipe}. 

Dataflow style architectures are an excellent fit to FPGAs as they can leverage FPGA programmability to specialize compute units to the needs of each layer within a specific network. However, with all layers now processing in parallel, the burden placed on the memory hierarchy increases significantly. To resolve this issue, previous implementations have stored network weights in on-chip buffers, which can provide very high bandwidth \cite{hpipe, aoc}. This allowed them to process smaller CNNs very efficiently but also precluded support for larger networks that did not fit on chip. One such example is HPIPE, a layer-pipelined CNN inference accelerator that leverages the reprogrammability of FPGAs to create custom hardware for each layer of a CNN. HPIPE leverages this per-layer customization in multiple ways, including having different compute units for traditional, depthwise and pointwise convolutions, sparse (zero-weight-skipping) convolutions~\cite{hpipe}, and convolutions mapped to AI optimized tensor blocks~\cite{hpipenx}. HPIPE also incorporates FPGA-specific physical optimizations such as as RAM fanout optimization and deep interconnect pipelining \cite{hpipe} to achieve high frequencies. On the networks that fit on-chip, HPIPE outperforms GPUs at low batch sizes. However, because it relies on on-chip memory, it is unable to process larger CNNs such as ResNets \cite{resnet_og_paper} unless they are retrained to have a high-degree of weight sparsity, which impacts their accuracy. In this work we augment HPIPE with off-chip weight storage to extend its high computational throughput to larger networks.

Some recent FPGAs are equipped with High-Bandwidth Memory (HBM), a DDR memory directly connected to them via interposers, allowing for higher bandwidth than previously possible \cite{jedec}. Since High-Bandwidth Memory was introduced, it has been widely adopted and studied across different types of devices such as CPUs, GPUs and FPGAs \cite{gpu_hbm, shuhai}. 
However, the way in which HBM should best be used to accelerate a dataflow-style architecture is still unclear. To our knowledge, this paper presents the first work to investigate how best to use HBM in a dataflow architecture specialized for CNN execution. We present the H2PIPE architecture and compiler-flow that intelligently offloads layers to HBM to minimize throttling of the compute pipeline and automatically generates all the necessary hardware. H2PIPE supports large networks and outperforms all prior work in CNN inference throughput.

The contributions of this paper are as follows:
\begin{itemize}
\item A characterization of HBM2 on the Stratix 10 NX FPGA and analysis of the implications on CNN dataflow architectures and the HPIPE architecture in particular. 
\item Enhancements to the HPIPE CNN accelerator architecture for efficient weight offload to HBM.
\item An analysis of theoretical peak throughput possible for ResNet-18, ResNet-50 and VGG-16 on the new H2PIPE architecture, with weights stored exclusively on HBM, partially on HBM, or with an infinite number of HBM stacks, showing that our implementations are close to upper-bound theoretical limits of performance.
\item An evaluation of the H2PIPE architecture on larger networks, showing that its throughput exceeds that of all previous FPGA solutions by at least 19.4x, 5.1x and 10.5x on ResNet-18, ResNet-50 and VGG-16 respectively at batch size 1.
\end{itemize}

\section{Background}
\subsection{Convolutional Neural Networks}
The convolutional neural networks discussed in this paper rely heavily on traditional 2D convolutions. In CNNs, 2D convolutions use a 4-dimensional weight tensor ($k_h \times k_w \times c_i \times c_o$) to operate on a 3-dimensional input tensor and produce a 3-dimensional output tensor. The convolution occurs across the width and height of the input tensor, with $k_h$ and $k_w$ representing the height and width of the filter, respectively. The input tensor has a depth of $c_i$, equal to the depth of the filters, while the output tensor has a depth of $c_o$, corresponding to the number of filters.

With most CNNs, as an image progresses through the network it shrinks along the height and width dimensions but sees an increase in the number of channels \cite{resnet_og_paper, vgg_og_paper, googlenet_og_paper, mv1_og_paper, mv2_og_paper, mv3_og_paper}, typically rising from 3 channels for an RGB input image to many channels at the input of the last layer (e.g. 2048 in the case of ResNet-50). 
\begin{figure}[h]
    \centering
    \includegraphics[width=\columnwidth]{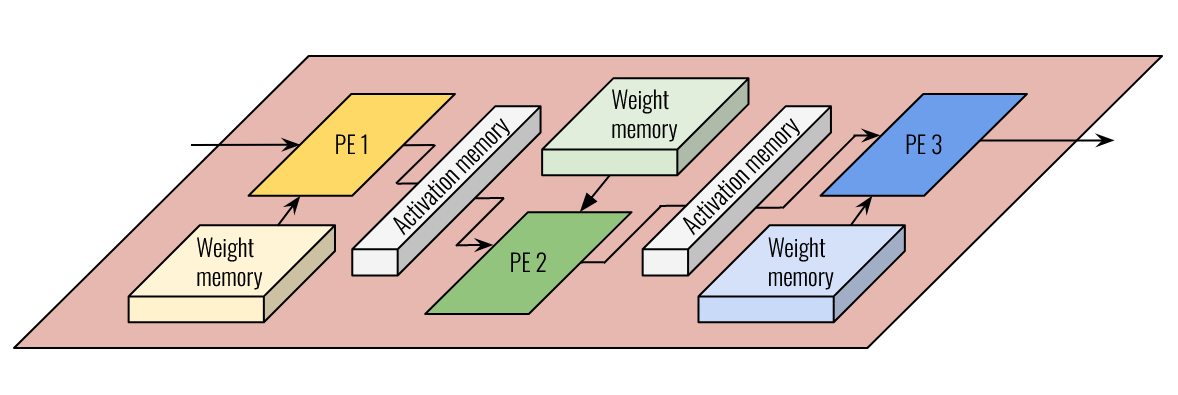}
    \caption{CNN dataflow architecture overview}
    \label{fig:dataflow}
\end{figure}
\subsection{CNN Dataflow Accelerators and Related Work}

 A simplified schematic of a dataflow architecture is shown in Figure \ref{fig:dataflow} where each processing element has its own weight memory connected to it, and with consecutive processing elements separated by activation memories. This allows all layers to process in parallel and for the processing element hardware to be specialized to each layer, creating a high throughput pipeline. The activation memories between layers only need to store a portion of each layer's computational results, generally only the number of lines needed by the next layer's convolutional kernel receptive field, making their on-chip storage attractive. 
 
 Most CNN dataflow accelerators also use only on-chip memory to store the weights \cite{aoc, hpipe, finn, xilinx_dataflow} due to the high weight bandwidth required\footnote{The work in \cite{xilinx_dataflow} can support storing the last layer of a network in off-chip memory such as DDR}. 
 The fpgaConvNet project~\cite{fpgaconvnet}, on the other hand, can use off-chip memory by processing a subset of CNN layers through the dataflow engine (with their weights on chip) before moving on to another subset of layers and transferring their weights from off-chip storage to on-chip before continuing processing. To reduce how often weight loads must be performed, they can use batch processing of multiple input images at a time, improving throughput at the cost of increased latency. 
 Follow-on work in \cite{satay} uses 8-bit weights and targets the YOLO family of networks for object detection. This work also reduced on-chip activation storage by buffering the skip connection data between layers in off-chip memory in order to free up more on-chip storage for the weights. 
The FINN architecture \cite{finn, xilinx_dataflow} relies on binary weights, reducing memory requirements and improving throughput at the cost of accuracy. 
The AoCStream architecture~\cite{aoc} is a layer-pipelined dataflow architecture targeting smaller FPGAs. It must fit all weights on chip and uses weight pruning to reduce weight storage.

HPIPE \cite{hpipe} originally used 16-bit precision, and also supports zero-weight skipping on weight-sparse networks to increase throughput and reduce on-chip weight memory. Later enhancements increased performance further by porting the HPIPE architecture to 8-bit precision and structuring the convolutions to leverage the AI-Optimized Tensor Blocks (AI-TBs) on the Stratix 10 NX FPGA~\cite{hpipenx}. 
HPIPE includes a compiler that chooses the amount of parallelism in each layer in order to allocate resources to create a balanced pipeline with throughput of all layers roughly matched. It always parallelizes computations across the entire width of activations, and chooses the number of input and output channels processed in parallel, $p_i^l$ and  $p_o^l$ for each layer $l$, to increase the throughput of layers that would otherwise bottleneck the computation.  
We build on the most recent (8-bit and AI-TB optimized) version of HPIPE as it has the highest reported FPGA performance on the networks it supports, including MobileNetV1~\cite{mv1_og_paper}, MobileNetV2~\cite{mv2_og_paper} and MobileNetV3~\cite{mv3_og_paper}.
Leveraging High-Bandwidth Memory will allow us to support larger networks like the ResNets \cite{resnet_og_paper} without compromising on precision or accuracy. 

There is relatively little work on CNN acceleration using High-Bandwidth memory on FPGAs.
In \cite{traininghbm}, Venkataramanaiah et al. use HBM for CNN training on FPGAs, demonstrating power gains over off-chip standard DDR3 memory.
The works in \cite{rwnn} and \cite{multicorehbm} use HBM for CNN inference on PE-style architectures. The former targets randomly wired neural networks while the latter evaluates its proposed system on VGG-16 networks.

\subsection{High-Bandwidth Memory}
\begin{figure}
    \centering
    \includegraphics[width=\columnwidth]{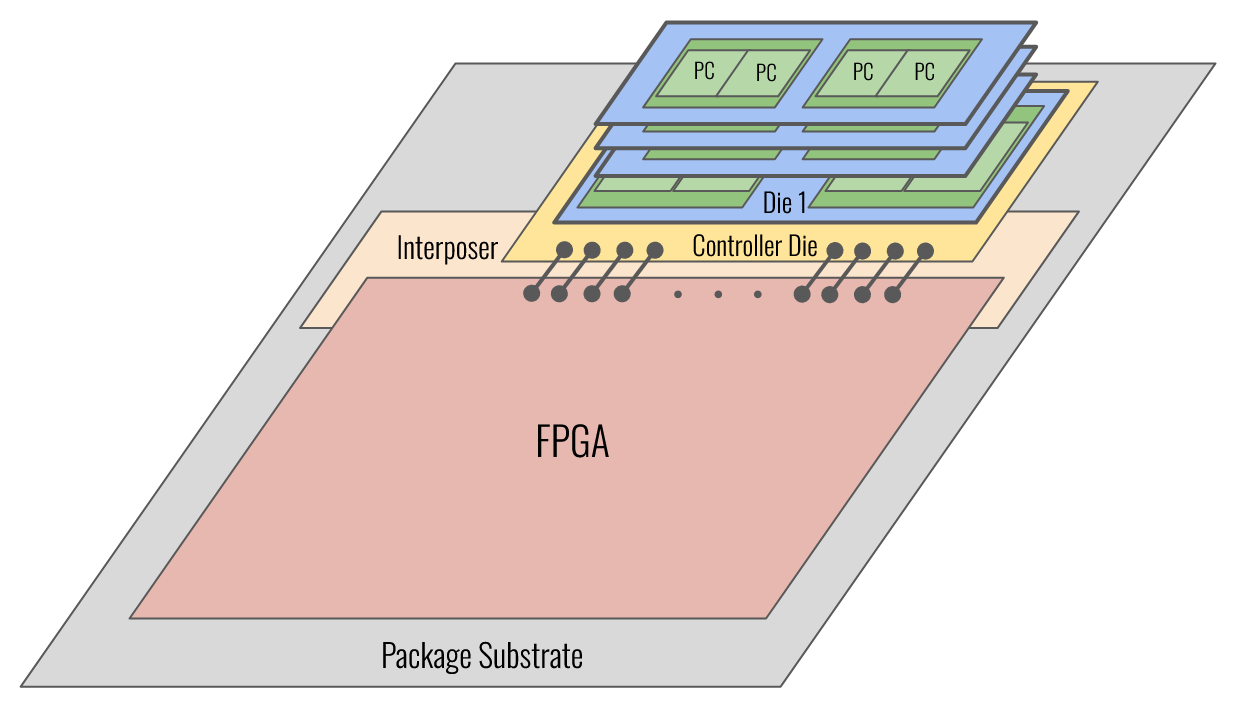}
    \caption{A 4-Hi HBM Stack connected to the top-side of an FPGA through an interposer}
    \label{fig:hbm_diagram}
\end{figure}
\begin{figure*}[th]
    \centering
    \begin{subfigure}[th]{0.5\textwidth}
        \centering
        \includegraphics[width=\columnwidth]{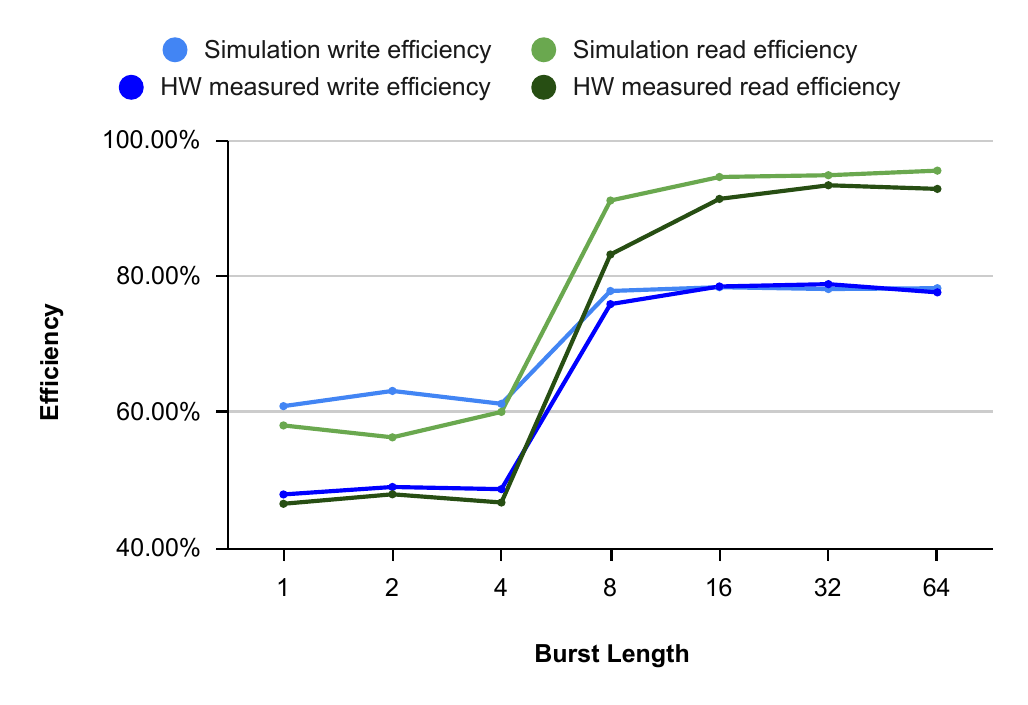}
        \caption{ }
        \label{fig:rd_w_efficiency}
    \end{subfigure}%
    ~
    \begin{subfigure}[th]{0.5\textwidth}
        \centering
        \includegraphics[width=\columnwidth]{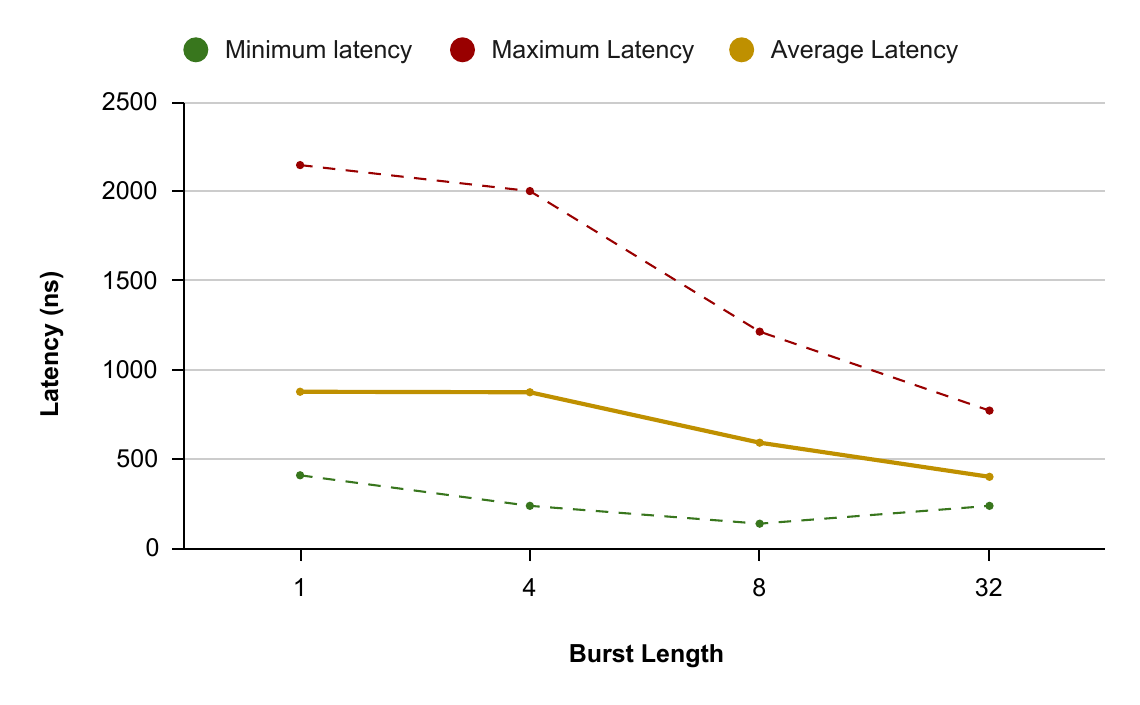}
        \caption{ }
        \label{fig:latency}
    \end{subfigure}
    \caption{(a) Simulated and hardware measured read and write efficiency of an HBM pseudo-channel as a function of burst length and (b)~Hardware measured read latency as a function of burst length when controller bandwidth is saturated}
\end{figure*}

High-Bandwidth Memory is a 3D-stacked DRAM for bandwidth-intensive applications~\cite{jedec}. We describe the HBM2 standard and refer to it as HBM throughout the rest of this paper. Each die in an HBM stack consists of 2 independent channels; each channel can be further split into 2 pseudo-channels (PCs) with separate controllers and unique address spaces but that share the channel's row and column command bus. Figure \ref{fig:hbm_diagram} shows a 4-Hi HBM stack that contains 16 Pseudo-Channels and can store up to 4 GB of data. There are 1024 data I/Os running at 800 MHz double data rate (i.e. 1.6 Gb/s) between the FPGA and HBM stack in the -2 FPGA speed grade we target for a total bandwidth of 204.8 GB/s. In order to support this large number of high-speed pins, the HBM stack is placed on a silicon interposer that reaches one edge of the FPGA; Stratix 10 NX has one HBM stack on the top of the device and another at the bottom for a total HBM bandwidth of 409.6 GB/s.
The FPGA logic communicates with a hardened HBM pseudo-channel controller over a 256-bit wide interface that runs at a maximum frequency of 400 MHz, matching the I/O data bandwidth at a more fabric-friendly frequency.

\section{Analysing High-Bandwidth Memory for H2PIPE}
HPIPE's dataflow architecture as drawn in Figure \ref{fig:dataflow} has two main consumers of block RAM: weights and activations, and a key design choice is which should be moved off-chip when on-chip RAM is insufficient. To inform this choice, we analyze  
the  read and write efficiencies and latencies achieved by HBM under address patterns that mimic HPIPE's address stream. 

\subsection{Characterizing HBM}
\label{sec:characterizing_hbm}

To characterize an HBM stack, we create an AXI traffic generator with selectable address patterns and burst lengths. We simulate the traffic generator in VCS MX using Quartus' HBM IP simulation script, and then validate the results in hardware on the top HBM stack of a Gidel Stratix 10 NX2100 board. For the graphs in Figure \ref{fig:rd_w_efficiency}, we issue reads and writes to random HBM addresses whenever the controller does not assert the back-pressure signal, saturating its bandwidth. We collect data over 10,000 write transactions first, followed by another 10,000 read transactions, and repeat for varying burst lengths. This best models the access pattern of multiple HPIPE layers either trying to read different weight kernels stored into a single pseudo-channel, or different layers trying to write activations at different addresses. We measure efficiency as the number of cycles in which the controller accepts a transaction over the total number of cycles. At $burst\_length \leq 4$, the efficiency of both reads and writes is slightly more than half of what it is at $burst\_length\geq8$. These findings correlate with the findings in \cite{shuhai} where smaller burst lengths yield around only 50\% the efficiency of larger burst lengths when accesses are non-sequential. Moreover, the simulation model coincides very closely with the hardware measurements at $burst\_length\geq8$, but is less accurate when $burst\_length\leq4$. We also note that the write efficiency peaks at around 15 percentage points lower than that of the read efficiency. 

We then measure the minimum, average and maximum latencies for reads as a function of the burst length when the read bandwidth is again saturated, as opposed to traditional idle latency measurements. This is because HPIPE is expected to keep controllers busy and at their bandwidth limit. As can be seen in Figure \ref{fig:latency}, latency decreases with burst length, reaching an average of 400 ns at burst length 32. When compared to the minimum latency, we find that the average latency when saturated is significantly higher: the minimum latency typically occurs at the very beginning of the measurements, when the controller isn't saturated with requests. 
Additionally, we find that when the accesses are less frequent and bandwidth is not saturated, or when reads happen to sequential addresses, the average latency remains below 450 ns irrespective of the burst length.

\subsection{H2PIPE Implications}
\label{sec:h2pipe_implications}
H2PIPE has two main sources of on-chip BRAM consumption: weights and activations. In Table \ref{tab:weight-act-ratio}, we compare the expected size of weight kernels and of the activation buffers in H2PIPE at minimum parallelism settings, including optimizations such as weight duplication and activation buffer duplication that improve $F_{\textrm{max}}$. In all compared networks, the activations represent less than 35\% of the memory requirements; this is partially due to the fact that HPIPE's layer-pipelined architecture means it is sufficient to store only a portion (a sliding window of lines on which computation is occuring) of the activations. Moreover, for the ResNets, activations represent less than 21\% of the memory requirements. For VGG-16, they represent even less than 2\% of the memory requirements. The Stratix 10 NX2100 device can only store 140 Mbits of data at a time in its BRAM, so offloading at least some of the weights to HBM is necessary to support ResNet-50 and VGG-16. 

\begin{table}
\centering
\begin{tabular}{||c | p{0.17\linewidth} | p{0.17\linewidth} |p{0.2\linewidth}||} 
 \hline 
 Model & \centering Weight Mem (Mb) & \centering Act Mem (Mb) & \centering Act Mem / Total Mem \cr
 \hline\hline
 MobileNetV1 & 35 & 11 & 24.1\% \\
 \hline
 MobileNetV2 & 29 & 15 & 34.5\% \\
 \hline
 MobileNetV3 & 32 & 12 & 27.0\% \\ 
 \hline
 ResNet-18 & 102 & 12 & 10.5\% \\ 
 \hline
 ResNet-50 & \cellcolor[gray]{0.8} 219 & 57 & 20.7\% \\
 \hline
 VGG-16 & \cellcolor[gray]{0.8} 1,204 & 14 & 1.1\% \\
 \hline
\end{tabular}
\caption{Memory required by HPIPE for several models. Shaded cells exceed the memory available on Stratix 10 NX2100}
\label{tab:weight-act-ratio}
\end{table}

Offloading either weights or activations to HBM will incur a cost in latency. As seen in Figure \ref{fig:latency}, the lowest average HBM read latency is 400 ns (which occurs at a burst length of 32). On MobileNetV2 for instance, choosing to offload every set of activations for each of the 53 convolutional layers, we are set to incur a latency increase of at least $53 \times 0.4 = 21$\textmu s. Considering that MobileNetV2 has a latency of 190 \textmu s on HPIPE, offloading activations would yield an increase of at least 11\% in latency. While it would seem that weights would have to pay the same price in order to be offloaded, HPIPE has an entirely deterministic sequence of weight reads. This means that by duplicating the control logic responsible for issuing reads, we can have it run hundreds of cycles in advance. At the output of HBM, buffers can hold the weights until they are actually required. To cover the latency of worst case reads at $burst\_length \geq 8$, buffers must hold data to keep the compute unit fed for at least 1214 ns, which translates to 364 cycles at 300 MHz. For that reason, we use 512 word deep FIFOs that can keep the pipeline running while covering even the worst-case HBM latency.

Tensor blocks in Stratix 10 NX compute three dot products of 10-element vectors each, per cycle. As the AI-TB has only 96 inputs from the programmable routing, it cannot provide sixty independent 8-bit inputs as that would require 480 total inputs per AI-TB. Instead, HPIPE uses a cascade chain to load input activations (30 values) into a set of ping-pong registers inside the AI-TB. Each cycle it broadcasts a 10 element vector of weights to all 3 dot-product units in order to compute a portion of three horizontally adjacent output channel elements. This means that each AI-TB requires 80-bits of weight data each cycle to feed its compute units. 
For more detail on how HPIPE organizes convolutions for high efficiency in AI-TBs, please refer to \cite{hpipenx}. Since each pseudo-channel can deliver 256 bits of data per cycle and each tensor chain requires 80 bits of weight data per cycle, we can feed at least 3 tensor chains with the output of each pseudo-channel. This will require interleaving read addresses from 3 tensor chains over the same pseudo-channel, creating a non-sequential read access pattern; however, it will achieve bandwidth at least as good as the random read accesses described in Section \ref{sec:characterizing_hbm}. The hardware measured read efficiency (dark green) line in Figure \ref{fig:rd_w_efficiency} shows HBM read bandwidth efficiency is much higher at a burst length of 8 than 4 or less, but from a burst length of 8 to 32 the efficiency increases more modestly from 83\% to 93\%. While larger burst lengths increase bandwidth efficiency, they necessitate larger on-chip burst-matching buffers, which reduce the space available for storing the weights of some layers on chip. The results of Figure~\ref{fig:rd_w_efficiency} indicate that burst lengths of 8, 16 and 32 are all interesting design points, and we explore the best burst length for various models further in Section \ref{sec:achieved_throughput}. 

\section{Architectural modifications}
\subsection{Efficient weight distribution from HBM}
\label{sec:sec_distribution}

\begin{figure*}[t]
    \centering
    \begin{subfigure}[t]{0.5\textwidth}
        \centering
        \includegraphics[width=\columnwidth]{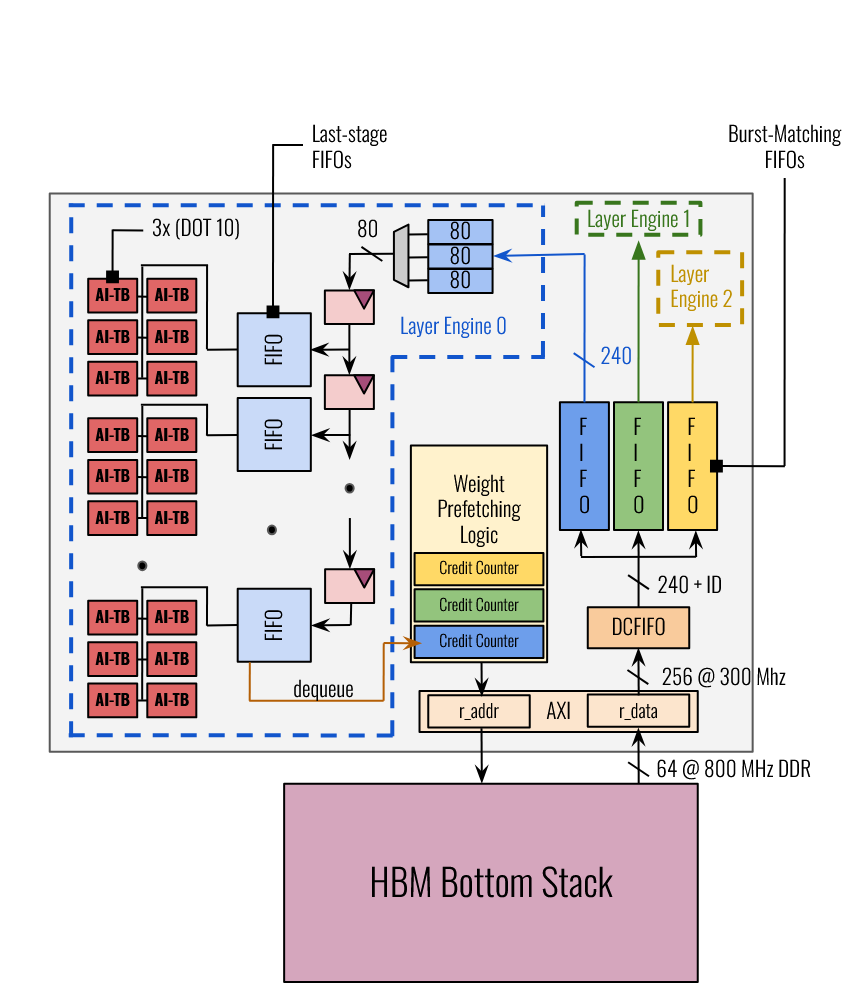}
        \caption{ }
        \label{fig:weight_distribution_network}
    \end{subfigure}%
    ~
    \begin{subfigure}[t]{0.5\textwidth}
        \centering
        \includegraphics[width=\columnwidth]{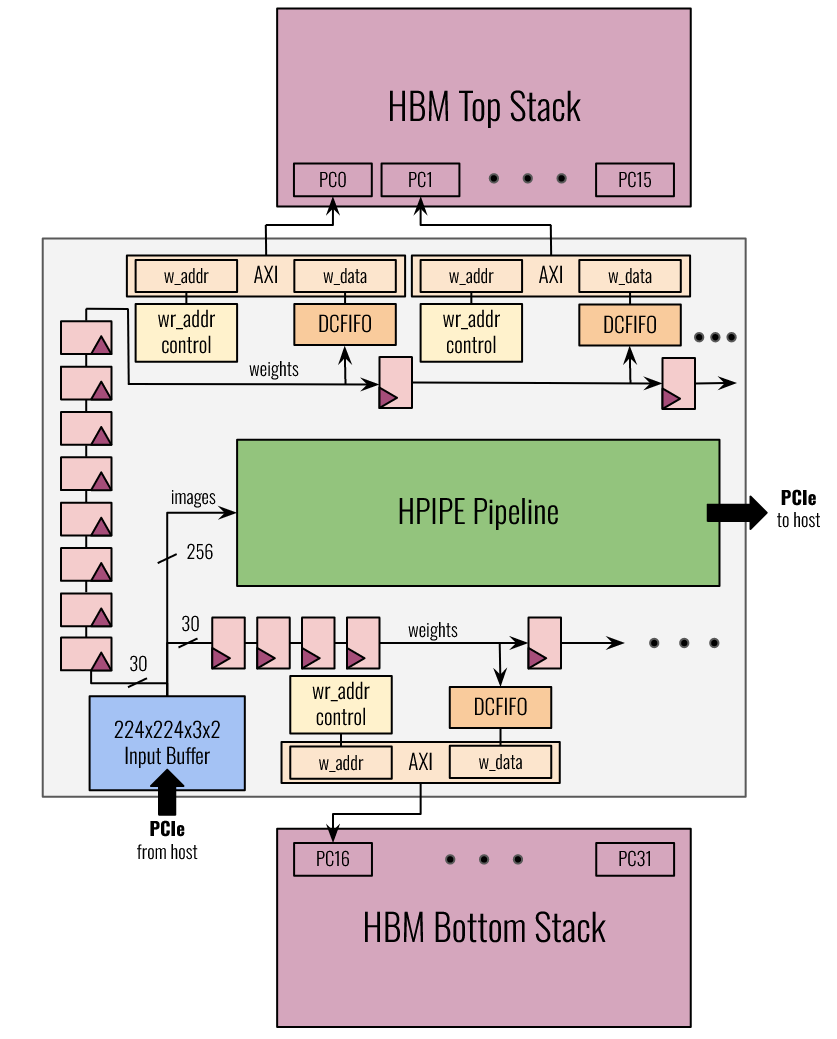}
        \caption{ }
        \label{fig:hbm_write_path}
    \end{subfigure}
    \caption{(a) Weight distribution network for the bottom HBM stack (b) Weight write path to both top and bottom HBM stacks}
\end{figure*}
With the HBM stacks on the Stratix 10 NX board located above and below the main die, we need to identify a good way to deliver the weights to the layer engines distributed all over the chip. Since the HPIPE compiler generates a custom accelerator for an input CNN, it is important that the distribution network generation automatically adapts to changes in the layer engines, and creates deeply pipelined and place-and-route-friendly logic to enable a high operating frequency for a wide range of accelelerators.
The overall structure of the weight distribution network is shown in Figure \ref{fig:weight_distribution_network}.  The weight prefetching logic, being detached from the main pipeline, operates entirely in the HBM clock domain. The data from the resulting reads is fed through a Dual-Clock FIFO (DCFIFO), and then sent to the burst-matching Single-Clock FIFO (SCFIFO) of the layer to which it belongs. The depth of the burst-matching FIFOs increases proportionally to the burst length we select, with each FIFO feeding a separate layer engine using different weights and requiring different address streams. From there, weights are serialized into a stream of 80-bit chunks that are distributed to the last-stage single-clock FIFOs. With HPIPE working across the entire width of an activation in parallel, weights are needed across all operating tensor chains simultaneously. To deliver the weights from HBM to all those tensor chains, the HPIPE compiler duplicates the last-stage FIFOs and arranges them in a pipelined daisy chain in order for them to be closer to groups of AI-optimized tensor blocks. We empirically determined that a group size of 6 is the best trade-off  between improving $F_{\textrm{max}}$ and minimizing the number of duplications; this means the pipeline depth of the distribution network and the number of duplicated FIFOs is automatically adjusted by the HPIPE compiler as the number of AI-TBs in a layer engine changes. These FIFOs are 80-bits wide and must be 512 words deep to cover the worst-case HBM read latency as discussed in Section \ref{sec:characterizing_hbm}. They consists of two M20Ks operating in 512 word x 40 bit wide mode. The FIFO control flow was initially implemented using almost\_full/ready signals, but we later redesigned it with a credit system to avoid deadlocks as described in Section \ref{sec:channel_sharing}. Figure \ref{fig:weight_distribution_network} shows the credit counters in the weight prefetching logic block; they are decremented when HBM read requests are issued and incremented by the `dequeue' signal from the layer engines when a set of weights has been fully consumed.

\subsection{Handling non-deterministic read latency}
When using on-chip weight buffers, the HPIPE pipeline never had to stall due to weights being unavailable as the access latency was deterministic. Section~\ref{sec:characterizing_hbm} showed that, so long as we use long bursts, HBM read efficiency is high (around 90\%), but it is still not 100\%. To maximize the number of layer engines that can have their weights moved to HBM, it is undesirable to overprovision the HBM bandwidth to a layer engine to ensure weights are always available; instead we accept that due to read efficiency being less than 100\% and/or occasional latency increases (e.g. during an HBM refresh cycle) that some on-chip weight distribution FIFOs may briefly empty. %
In order to avoid erroneous computations, there are two ways to handle empty weight FIFOs. Computations can be run in batches, with a batch only starting when enough weights are present to last the entire batch. Alternatively, computations can run continuously, with a `freeze' signal available to stall the tensor chain and all related elements when FIFOs are about to run out of weights. The second option can be efficiently implemented  and minimizes stalling, so we choose it.
We use the last-stage FIFO `almost\_empty' to freeze the computations. Connecting this signal directly to the `clk\_enable' port of all registers involved in the computation would be the straightforward way to stall the computation, but is not desirable in Stratix 10. First, the freeze signal would have very high fanout, leading to a long interconnect delay that can limit $F_{\textrm{max}}$. Second, Stratix 10 has optional registers in each routing wire driver~\cite{stratix10}, allowing very deep pipelining that HPIPE exploits, but these registers are simple and do not have a clock enable. Consequently, freezing all the logic with a clock enable would prevent use of interconnect registers, making both re-timing of the circuitry less effective and placement of the pipeline registers more difficult. We instead connect the freeze signal only to the `clk\_enable' port of the AI-TBs, and integrate it as part of the control logic of input activation memories and output buffers.  This allows us to maintain the high $F_{\textrm{max}}$ of the H2PIPE pipeline with little overhead. 

\subsection{Creating a resource-efficient HBM write path}
At boot-up, we send weights from the host over PCIe to the H2PIPE accelerator for it to store into HBM. Because this operation is done only once and is not timing critical, we wish to allocate as few resources as possible to it. The write path can be found in Figure \ref{fig:hbm_write_path}. To avoid creating a new on-chip buffer (wasting BRAM) and complicating the datapath from the PCIe interface, we re-use the image input buffer of size $224\times224\times3\times2$ and its corresponding datapath. We upgrade the H2PIPE compiler to generate binary files containing the weight data formatted as input images. The simple way to write the weight data to HBM would be to create a 256 bit wide datapath from the input buffer to each DCFIFO in order to match the 256-bit width of the HBM pseudo-channel data interfaces. However, as this wide bus traverses a long distance from the PCIe block and input buffer to both the top and bottom HBM stacks as seen in \ref{fig:hbm_write_path}, and is heavily pipelined to meet timing, it consumes a large number of registers and significant wiring. Since weights are written only once when the accelerator begins operation, we instead narrow the write path that leads to both HBM stacks, deserializing it only right before the AXI controller. Users can choose the write path width via an input parameter to the H2PIPE compiler; we set a default width of 30 bits, which saves over 3000 registers compared to a straightfoward 256-bit wide interface. 

\section{Pseudo-channel sharing and layer selection}
\subsection{Pseudo-channel sharing and deadlock avoidance}
\label{sec:channel_sharing}
\begin{figure}
    \centering
    \includegraphics[width=\columnwidth]{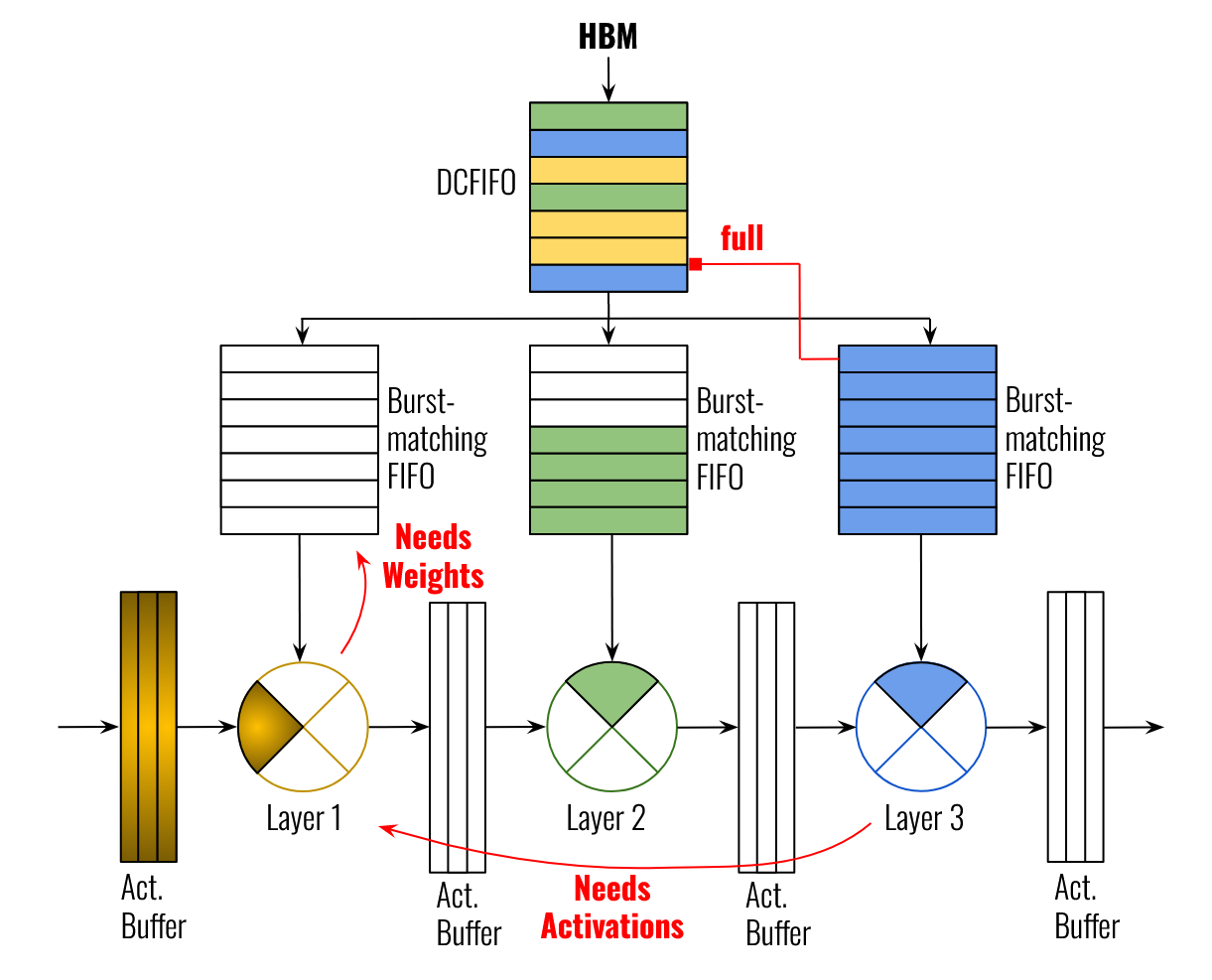}
    \caption{Deadlock scenario in a valid/ready latency-insensitive paradigm}
    \label{fig:deadlock}
\end{figure}

H2PIPE uses a latency-insensitive design between layers both to ensure correctness when operation latencies like HBM reads are non-deterministic and to allow the H2PIPE compiler to choose how many pipeline registers to add on interconnect paths to achieve high $F_{\textrm{max}}$.
One way to implement the latency-insensitive weight distribution network described in section \ref{sec:sec_distribution} is via a valid/ready paradigm as described in \cite{carloni}. However, with each HBM to fabric DCFIFO feeding multiple burst-matching FIFOs, we find scenarios where one of the burst-matching FIFOs being full can lead to all of them being starved or worse, deadlocked. One such example is illustrated in Figure \ref{fig:deadlock}, where three consecutive layers share an HBM to fabric DCFIFO. The head of the DCFIFO is a word meant for layer 3, but the layer 3 burst-matching FIFO is full. In order for layer 3 to create space in its FIFO, it needs to receive activations from layer 2, that in turn needs to receive them from layer 1. However, layer 1 has no weights available due to the head of line blocking with all of layer 1's weights stuck behind the blue word in the HBM to fabric DCFIFO. This deadlocked state can be reached in many ways with a ready/valid paradigm; one example is at start up when the first layer is operating but the two consecutive layers are waiting on activations. To avoid these many deadlock cases, H2PIPE uses a credit-based latency-insensitive design style instead of the ready/valid protocol used in the original HPIPE. We add credit counters to the weight prefetching logic as drawn in Figure \ref{fig:weight_distribution_network}, ensuring no extra weights are sent to the FIFOs downstream, thereby guaranteeing that we never run into head of line blocking and deadlocks. %

\subsection{Choosing layers and physical constraints}
\label{sec:choosing_layers}
H2PIPE supports storing weights on-chip or in High-Bandwidth Memory. On-chip memory has high bandwidth but limited capacity, while HBM has high capacity but less bandwidth. Accordingly, the best candidates for offloading weights to HBM are layers with large weight storage needs but lower bandwidth requirements. Equation \ref{eq:score} scores the desirability of moving the weights of layer $l$ to HBM.

\begin{equation}
\label{eq:score}
    \textrm{score}_l = \frac{ \left(\left \lceil \frac{k^l_h \cdot k^l_w \cdot c^l_i \cdot c^l_o \cdot 8}{20480} \right \rceil - 2 \right) \cdot \left \lceil\frac{\textrm{output\_width}_l}{18} \right \rceil}{p^l_i \cdot p^l_o \cdot 80} 
\end{equation}

$p_i$ and $p_o$ are the degrees of input channel and output channel parallelism. $k_h$ and $k_w$ are the kernel height and width, $c_i$ and $c_o$ are the number of input and output channels respectively, and output\_width is the width of the activation tensor output by layer $l$. The score metric effectively divides the amount of M20Ks that will be saved by offloading each kernel (taking into account the fact that 2 M20Ks will replace every on-chip weight memory, and including effects such as weight memory duplication), by the bandwidth the corresponding layer will require. The bandwidth required is $p_i \cdot p_o \cdot 80$, where each degree of parallelism processes another 80 bits of weights and data in parallel. Algorithm \ref{alg:offload} selects which layers are offloaded to HBM based on the score metric defined above. It offloads the weights of layers most suitable for HBM until it runs out of bandwidth to allocate. Once we have decided which layers should have their weights stored in HBM, we need to choose which of the 32 HBM pseudo-channels each weight-offloaded layer should connect to. While the function of each pseudo-channel is the same, they are physically distributed across most of the top and most of the bottom of the FPGA, so this allocation has a significant impact on placement and routing. We find that a clock-wise assignment works best as it roughly matches the typical layout of the H2PIPE dataflow pipeline. More specifically, using the numbering convention depicted in Figure \ref{fig:hbm_write_path}, we assign weight-offloaded layers ordered from the CNN input to output to pseudo-channels ordered from $0 \rightarrow 15$ followed by $31 \rightarrow 16$.

\algnewcommand{\LineComment}[1]{\State \(\triangleright\) #1}
\algnewcommand{\ActualComment}[1]{\State \(//\) #1}
\algrenewcommand\algorithmicrequire{\textbf{Input:}}
\algrenewcommand\algorithmicensure{\textbf{Output:}}
\begin{algorithm}
\caption{Layer offloading algorithm}\label{alg:offload}
\begin{algorithmic}
\Require{$\texttt{n\_pc}, \texttt{\{p}_\texttt{i}^l\texttt{\}}_{l \in \texttt{0}, \cdots, \texttt{L-1}}, \texttt{\{p}_\texttt{o}^l\texttt{\}}_{l \in \texttt{0}, \cdots, \texttt{L-1}}$}
\Ensure{$\texttt{\{offload}_l\texttt{\}}_{l \in \texttt{0}, \cdots, \texttt{L-1}}$}
\ActualComment $\texttt{n\_pc}$ is the number of Pseudo-channels
\For{$l \in \texttt{0,} \cdots \texttt{,L-1}$}
    \LineComment Compute $\texttt{score}_l$ using Equation (l)
    \State $\texttt{offload}_l \gets \texttt{False}$    
\EndFor
\State $\texttt{\{order}_l\texttt{\}}_{l \in \texttt{0}, \cdots, \texttt{L-1}}$ $\gets$ layer numbers sorted by score
\State $\texttt{free\_BW} \gets \texttt{n\_pc} \times \texttt{3}$
\State $idx \gets \texttt{0}$
\While{$\texttt{free\_BW} \neq \texttt{0} \And idx < \texttt{L}$}
\State $l$ $\gets \texttt{order}_{idx}$
\If{$\texttt{p}_\texttt{i}^{l} \cdot \texttt{p}_\texttt{o}^{l} <= \texttt{free\_BW}$}

    \State $\texttt{offload}_l \gets \texttt{True}$
    \State $\texttt{free\_BW} \gets \texttt{free\_BW} - \texttt{p}_\texttt{i}^{l} \cdot \texttt{p}_\texttt{o}^{l}$
\EndIf
\State $idx \gets idx + \texttt{1}$
\EndWhile
\end{algorithmic}
\end{algorithm}

\section{Performance Results}
In this section, we describe data collected on the Gidel Stratix 10 NX2100 board with two 4-Hi HBM stacks.
We present results on ResNet-18, ResNet-50 and VGG-16 over the full ImageNet validation set (50,000 images). We first summarize the hardware results, then compare to two theoretical upper bounds on throughput, and finally compare throughput and latency to other state-of-the-art FPGA-based CNN accelerators.

\subsection{Achieved Throughput}
\label{sec:achieved_throughput}
We first measure the performance of ResNet-18, ResNet-50 and VGG-16 on H2PIPE in hardware storing all weights on HBM and using $burst\_length=8$. As shown by the dark blue bar in Figure~\ref{fig:theoretical}, ResNet-18 achieves 1811 im/s while ResNet-50 achieves 748 im/s and VGG-16 achieves 430 im/s. By using as many on-chip weight buffers as possible and using Algorithm~\ref{alg:offload} to choose which layers should be placed in HBM, we can keep more compute units busy and achieve higher throughput. The dark green bars in Figure~\ref{fig:theoretical} show the performance of this hybrid memory system: throughput increases significantly, to 4174 im/s, 1004 im/s and 545 im/s for ResNet-18, ResNet-50 and VGG-16 respectively. While all three networks benefit from the hybrid memory system, ResNet-18 benefits more as its smaller number of activation and weight buffers mean a larger fraction of the weights can be stored on chip.

As discussed in Section~\ref{sec:h2pipe_implications},  the burst length for HBM reads is another important design parameter. Table \ref{tab:throughput_burst} summarizes the throughput achieved with the hybrid memory system for both ResNet-18 and ResNet-50 with different burst lengths. Burst lengths of 8 and 16 achieve the same throughput on ResNet-18 as the pipeline bottlenecking layer in this network is using on-chip memory, and hence the HBM efficiency difference between these burst lengths has no throughput impact. ResNet-50 on the other hand sees its performance change with burst length, indicating its bottlenecking layer is stored on HBM. Increasing the burst length from 8 to 32 yields a 2\% increase in throughput at the price of logic utilization increasing from 78\% to 81\%. Our conclusion with regards to the optimal burst length is clear: on networks where the bottlenecking layer is stored on-chip (e.g. ResNet-18), a burst length of 8 will allow us to save on logic utilization. Conversely, on networks where the bottlenecking layer will have its weights stored in HBM (e.g. ResNet-50, VGG-16), H2PIPE uses a burst length of 32 to extract some additional performance at the cost of some extra logic. 

The Top-1/Top-5 accuracies for ResNet-18, ResNet-50 and VGG-16 are 62.4\%/84.6\%, 71.5\%/90.5\% and 65.8\%/87.3\% respectively. 
The 8-bit networks were trained with a simple 8-bit quantization approach, using int8 fine-tuning on models trained in fp32. Our efforts focus on improving the throughput of the system, and a better training scheme would result in higher accuracy, as int8 ResNets have been demonstrated to be able to reach comparable precision to fp32 implementations with more sophisticated quantization-aware training methods \cite{8bit_int_accuracy} .

\begin{table}
\centering
\begin{tabular}{| c | c | c | c |} 
 \hline 
 Model & Burst Length & Logic Utilization & Throughput (im/s) \\
 \hline
 \hline
 \multirow{2}{*}{ResNet-18} & 8 & 67\% & 4174 \\ 
 \cline{2-4}
 & 16 & 67\% & 4174 \\
 \hline
 \multirow{3}{*}{ResNet-50} & 8 & 78\% & 984 \\
 \cline{2-4}
 & 16 & 79\% & 988 \\
 \cline{2-4}
  & 32 & 81\% & 1004 \\ 
 \hline
\end{tabular}
\caption{Throughput of H2PIPE as a function of burst length}
\label{tab:throughput_burst}
\end{table}

\subsection{Comparisons to theoretical upper bounds}
\begin{figure}
    \centering
    \includegraphics[width=\columnwidth]{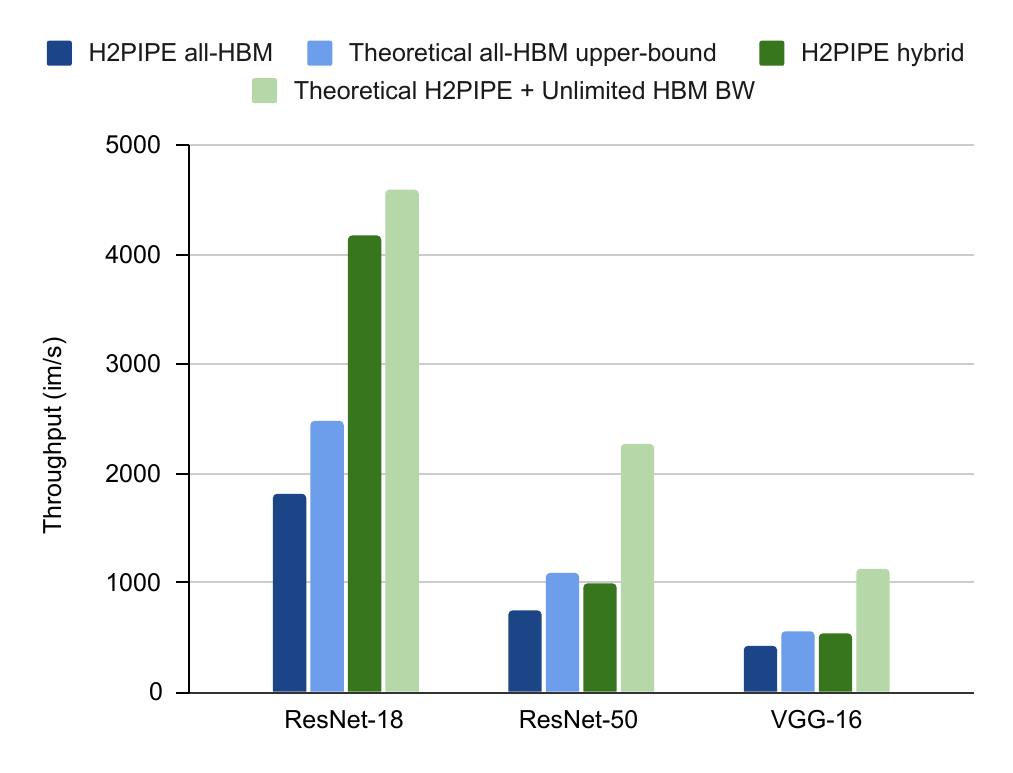}
    \caption{Hardware all-HBM and hybrid throughput numbers, compared with all-HBM theoretical upper-bound and unlimited-HBM-bandwidth bound}
    \label{fig:theoretical}
\end{figure}

\definecolor{Gray}{gray}{0.9}
\begin{table*}[htbp]
\begin{center}
\makebox[\textwidth][c]{
\resizebox{1.05\textwidth}{!}{
    \begin{tabular}{m{6em} |>{\centering\arraybackslash}m{1.4cm} >{\centering\arraybackslash}m{1.4cm} >{\columncolor{Gray}\centering\arraybackslash}m{1.55cm} |>{\centering\arraybackslash}m{1.4cm}
    >{\centering\arraybackslash}m{1.4cm}
    >{\centering\arraybackslash}m{1.55cm}
    >{\centering\arraybackslash}m{1.4cm} 
    >{\centering\arraybackslash}m{1.4cm}
    >{\columncolor{Gray}\centering\arraybackslash}m{1.55cm}
    |>{\centering\arraybackslash}m{1.4cm}
    >{\centering\arraybackslash}m{1.4cm}
    >{\centering\arraybackslash}m{1.6cm}
    >{\columncolor{Gray}\centering\arraybackslash}m{1.55cm}}
      \toprule[3pt]
           & Venieris et al. \cite{unzipfpga} & FILM-QNN \cite{filmqnn} & \textbf{H2PIPE (ours)} & Venieris et al. \cite{unzipfpga} & Liu et al. \cite{bestresnet50} & DNNVM \cite{dnnvm} & FTDL \cite{ftdl} & BNN-PYNQ \cite{xilinx_dataflow} \cite{bnn_batchone} & \textbf{H2PIPE (Ours)} & fpgaconvnet \cite{fpgaconvnet_vgg} & Ma et al. \cite{best_vgg16} & Nguyen and Nakashima~\cite{multicorehbm} & \textbf{H2PIPE (Ours)}\\
    \midrule[2pt]
    Device & Z7045 & ZC102 & Stratix 10 NX & ZU7EV & Arria 10 GX & ZU9 & VU125 & Alveo U250 & Stratix 10 NX & Z7045 & Stratix 10 GX & Alveo U280 & Stratix 10 NX \\ \midrule[0.25pt]
    Technology & 28nm & 16nm & 14nm & 16nm & 20nm & 16nm & 20nm & 16nm & 14nm & 28nm & 14nm & 16nm & 14nm \\ \midrule[0.25pt]
    Device BRAM (Mb) & 19.2 & 32.1 &  140 & 38 & 65.7  &  164 & 32.1 & 432 & 140 & 19.2 & 229 & 357 & 140 \\ \midrule[0.25pt]
    DSPs/Tensor Blocks  & 900 & 2,520 &  3,960 & 1,728 & 1,518 & 2,520 & 1,200 & 11,508 & 3,960 & 900 & 5,780 & 9,024 & 3,960 \\\midrule[2pt]
    Logic \ \ \ \ \ \ \ \ \ \ \ Utilization & -- & 66\% &  67\%  & -- & 71\% & -- & 75\% & 77\% & 81\% & -- & 50\% & 55\% & 64\% \\ \midrule[0.25pt]
    BRAM \ \ \ \ \ \ \ \ \ Utilization & -- & 48\% &  98\%  & -- & 86\% & 86\% & 37\% & 97\% & 98\% & -- & 21\% & 92\% & 91\% \\ \midrule[0.25pt]
    Used DSPs & 100\% & 83\% & 51\% & 100\% & 97\% & 61\% & 100\% & 14\% & 33\% & 95\% & 71\% & 96\% & 40\% \\ \midrule[0.25pt]    
    Frequency (MHz) & 150  & 150 & 300 & 200 & 200 & 500 & 650 & 195 & 300 & 125 & 300 & 250 & 300 \\\midrule[2pt]
    Network & ResNet-18 & ResNet-18 &  ResNet-18 & ResNet-50 & ResNet-50 & ResNet-50 & ResNet-50 & ResNet-50 & ResNet-50 & VGG-16 & VGG-16 & VGG-16 & VGG-16 \\ \midrule[0.25pt]
    Precision & 16-Bit & 95\% 4-Bit + 5\% 8-Bit &  8-Bit & 16-Bit & 8-Bit &  8-bit & 16-Bit & 1-Bit & 8-Bit & 16-Bit & 8-Bit & 16-Bit & 8-Bit \\ \midrule[0.25pt]
    Format & Fixed & Mixed & Fixed & Fixed & Fixed & Fixed & Fixed & Fixed & Fixed & Fixed & Fixed & Fixed & Fixed \\\midrule[2pt]
    Throughput \footnotesize{($B$=$1$, $im/s$)}& 59.7 & 214.8 & 4,174 & 71.7 & 197.2 & 88.3 & 151.2 & 527 & 1,004 & 4.0 & 51.8 & 29.5 \footnotemark[1] & 545\\ \midrule[0.25pt]
    Latency \ \ \ \ \footnotesize{($B=1$, $ms$)}& 16.75 & -- &  1.01 & 13.95 & 5.07 &  -- & 6.61 & 1.90 & 9.48 & 249.5 & 19.29 & 33.92 & 9.76 \\ \midrule[0.25pt]
    GOPs \footnotesize{($B$=$1$)} & 236 & 779 & 15,109 & 603 & 1,519 & 680 & 1,164 & 3,567 & 7,731 & 156 & 1,605 & 913 \footnotemark[1] & 16,873 \\
    \bottomrule[3pt]%
    \end{tabular}
}
}
\end{center}
\footnotesize{\footnotemark[1] Throughput and GOPS of \cite{multicorehbm} reported with a favorable batch size of 128.}
\caption{Comparison to other high performance CNN accelerators on FPGAs.}
\label{tab:accelerator_comparisons}
\end{table*}

To provide further context on the effectiveness of our memory system, we compare our hardware results to two unachievable upper bounds. First, we compare the H2PIPE all-HBM hardware results to a throughput limit: peak HBM bandwidth, divided by the total weight memory traffic required to process an image in the H2PIPE pipeline.
In computing the peak HBM bandwidth we assume an unrealizable 100\% read efficiency, but only use 31 of the 32 pseudo-channels because using PC16 (see Figure \ref{fig:hbm_write_path})  creates timing closure issues due to its proximity to the secure device manager found at the bottom of the chip. Choi et al \cite{hbm_interconnect} also decide to leave out two pseudo-channels when trying to extract the maximum performance of HBM for similar reasons. Furthermore, due to tensor chains requiring multiples of 80 bits of data per cycle, we only use 240 bits of 256 bits of data that HBM can provide. With the core logic running at 300 MHz, this translates to a maximum available HBM bandwidth of 279 GB/s. HPIPE parallelizes computations across the width of input activations, meaning it only reloads the kernels for each line in the image. Using that observation, the required weight memory traffic for a network that is run by H2PIPE can be found to be
\begin{equation}
MT_{\textrm{required}} = \sum_{i=1}^{\#\textrm{layers}} k_h * k_w * c_i * c_o * \textrm{output\_height}  
\end{equation} With the required bandwidth for each network and the effectively available bandwidth of 279 GB/s, we calculate a throughput upper-bound that a completely optimal implementation would have achieved with perfect HBM efficiency and perfect layer to pseudo-channel assignment. The results are shown as the light blue bars in Figure \ref{fig:theoretical}. Comparing to the hardware all-HBM results (dark blue bars), we find that H2PIPE's performance ranges between 68\% (ResNet-50) and 78\% (VGG-16) of the theoretical upper-bound, indicating that our implementation leverages HBM efficiently. 
For ResNet-18, the hybrid approach as seen in the dark green bar achieves almost double the throughput of this theoretical all-HBM upper bound, highlighting the importance of combining on-chip and HBM memory for weight buffers.

To prove that our implementation would scale even as more HBM stacks become available, we also run simulations assuming unlimited HBM bandwidth (i.e. more than 2 HBM stacks) and increase DSP count until 85\% of logic or DSP utilization is reached. The resulting throughput numbers are drawn in the fourth (light green) vertical bar in Figure \ref{fig:theoretical}. ResNet-18 would not benefit significantly from the addition of HBM stacks as its bandwidth requirements are almost completely satisfied by the available on-chip memory and two HBM stacks in the hybrid case; it is limited mostly by compute resources. On the other hand, ResNet-50 and VGG-16 could achieve up to 2.27x and 2.08x increases in throughput respectively before becoming limited by compute resources.

\subsection{Comparison to prior work}
Table \ref{tab:accelerator_comparisons} compares the performance of H2PIPE to several state-of-the-art FPGA CNN accelerators on ResNet-18, ResNet-50 and VGG-16. For ResNet-18, we outperform the work in \cite{unzipfpga} by 69.9x, although we do use a more recent process technology (14 nm instead of 28 nm) and 8 bits instead of 16. We also outperform the work in \cite{filmqnn} that uses a similar process technology to our work and a lower precision (mixed 4- and 8-bit precision vs. our 8-bit precision) by 19.4x. On ResNet-50, our accelerator also outperforms other 8- and 16-bit accelerators by at least 5.1x, with speed-ups of up to 14x. H2PIPE achieves a high frequency of 300 MHz, but \cite{dnnvm} and \cite{ftdl} achieve even higher respective frequencies of 500 and 650 MHz. Despite that, we still outperform them by 11.4x and 6.64x respectively, showing that a combination of HBM and a layer-pipelined dataflow architecture can deliver very high performance. The layer pipeling of H2PIPE helps throughput, but has some cost in latency: \cite{bestresnet50} and \cite{ftdl} obtain 47\% and 30\% lower latency than H2PIPE respectively, but have 5.1x and 6.64x less throughput. Finally, even when compared to the binarized network in \cite{xilinx_dataflow} at batch=1, H2PIPE has almost double the throughput, while doing far more work on a smaller device. We note that the binarized accelerator is the only other ResNet-50 accelerator in the table that employs a dataflow architecture; it is able to fit all weights on-chip thanks to binary quantization. On VGG-16, we compare against the only other HBM-enabled CNN inference accelerator in~\cite{multicorehbm} which uses 16-bit precision but a larger device.
We achieve 18.5x their throughput and 3.48x lower latency. Compared to the highest performance prior VGG-16 work~\cite{best_vgg16} (which does not use HBM), we achieve a 10.5x speed-up while also having half the latency on a smaller device.

\section{Conclusion}
We augmented the HPIPE architecture to overcome a major challenge of most dataflow CNN accelerators -- supporting CNNs too large to fit entirely in on-chip memory while maintaining layer specialized hardware and high throughput. We found that HBM memory could be best leveraged by offloading kernel weights, while keeping activations on chip, due to the larger weight kernel memory consumption and the fact that the high HBM latency could be covered by creating streaming weight buffers and pre-fetching weights well before the compute units need them. Sharing an HBM pseudo-channel between multiple layer engines improves usage of HBM bandwidth but introduces the possibility of deadlocks. A credit-based latency-insensitive design style solves deadlock issues while maintaining correctness as the H2PIPE compiler composes layer engines into an accelerator and automatically pipelines interconnect. The H2PIPE compiler also automatically constructs a hybrid memory system in which layers with lower weight bandwidth to size ratios are placed in HBM while others are kept on chip. Hardware measurements on the 50,000 ImageNet validation dataset show H2PIPE has 19.4x, 5.1x and 10.5x higher throughput on ResNet-18, ResNet-50 and VGG-16 than the best prior work using comparable precision. 

Future work could extract even more bandwidth out of HBM when using H2PIPE by separating the HBM interface clock from the layer engine clock (even when the layer engines do not run at 400 MHz), at the cost of more on-chip buffering. Other future work could leverage neural architecture search (NAS) to optimize over the very large space of accelerators that H2PIPE can create. In particular, H2PIPE can take advantage of layers with large kernels and low parallelism settings by storing their weights on HBM, or small layers with high parallelism settings by storing their weights on-chip. 
Such trade-offs can be explored and exploited by NAS to produce even higher performance inference.

\section{Acknowledgements}
This work was funded by the Intel/VMWare Crossroads 3D-FPGA Academic Research Centre and NSERC. We thank the anonymous reviewers for their helpful comments.

\bibliographystyle{./bibliography/IEEEtran}
\bibliography{ref.bib}

\begin{thebibliography}{10}
\providecommand{\url}[1]{#1}
\csname url@samestyle\endcsname
\providecommand{\newblock}{\relax}
\providecommand{\bibinfo}[2]{#2}
\providecommand{\BIBentrySTDinterwordspacing}{\spaceskip=0pt\relax}
\providecommand{\BIBentryALTinterwordstretchfactor}{4}
\providecommand{\BIBentryALTinterwordspacing}{\spaceskip=\fontdimen2\font plus
\BIBentryALTinterwordstretchfactor\fontdimen3\font minus \fontdimen4\font\relax}
\providecommand{\BIBforeignlanguage}[2]{{%
\expandafter\ifx\csname l@#1\endcsname\relax
\typeout{** WARNING: IEEEtran.bst: No hyphenation pattern has been}%
\typeout{** loaded for the language `#1'. Using the pattern for}%
\typeout{** the default language instead.}%
\else
\language=\csname l@#1\endcsname
\fi
#2}}
\providecommand{\BIBdecl}{\relax}
\BIBdecl

\bibitem{beyond}
A.~Boutros, E.~Nurvitadhi, R.~Ma, S.~Gribok, Z.~Zhao, J.~C. Hoe, V.~Betz, and M.~Langhammer, ``Beyond peak performance: Comparing the real performance of {AI}-optimized {FPGA}s and {GPU}s,'' in \emph{2020 International Conference on Field-Programmable Technology (ICFPT)}, 2020, pp. 10--19.

\bibitem{survey}
\BIBentryALTinterwordspacing
K.~Abdelouahab, M.~Pelcat, J.~S{\'{e}}rot, and F.~Berry, ``Accelerating {CNN} inference on {FPGA}s: {A} survey,'' \emph{CoRR}, vol. abs/1806.01683, 2018. [Online]. Available: \url{http://arxiv.org/abs/1806.01683}
\BIBentrySTDinterwordspacing

\bibitem{dla}
\BIBentryALTinterwordspacing
M.~S. Abdelfattah, D.~Han, A.~Bitar, R.~DiCecco, S.~O'Connell, N.~Shanker, J.~Chu, I.~Prins, J.~Fender, A.~C. Ling, and G.~R. Chiu, ``{DLA:} compiler and {FPGA} overlay for neural network inference acceleration,'' \emph{CoRR}, vol. abs/1807.06434, 2018. [Online]. Available: \url{http://arxiv.org/abs/1807.06434}
\BIBentrySTDinterwordspacing

\bibitem{xilinx_dataflow}
\BIBentryALTinterwordspacing
L.~Petrica, T.~Alonso, M.~Kroes, N.~J. Fraser, S.~Cotofana, and M.~Blott, ``Memory-efficient dataflow inference for deep {CNN}s on {FPGA},'' \emph{CoRR}, vol. abs/2011.07317, 2020. [Online]. Available: \url{https://arxiv.org/abs/2011.07317}
\BIBentrySTDinterwordspacing

\bibitem{hpipe}
M.~Hall and V.~Betz, ``From tensorflow graphs to {LUTs} and wires: Automated sparse and physically aware {CNN} hardware generation,'' in \emph{2020 International Conference on Field-Programmable Technology (ICFPT)}, 2020, pp. 56--65.

\bibitem{aoc}
\BIBentryALTinterwordspacing
H.-J. Kang, ``{AoCStream}: All-on-chip {CNN} accelerator with stream-based line-buffer architecture,'' in \emph{Proceedings of the 2023 ACM/SIGDA International Symposium on Field Programmable Gate Arrays}, ser. FPGA '23.\hskip 1em plus 0.5em minus 0.4em\relax New York, NY, USA: Association for Computing Machinery, 2023, p.~48. [Online]. Available: \url{https://doi.org/10.1145/3543622.3573141}
\BIBentrySTDinterwordspacing

\bibitem{hpipenx}
M.~Stan, M.~Hall, M.~Ibrahim, and V.~Betz, ``{HPIPE NX}: Boosting {CNN} inference acceleration performance with {AI}-optimized {FPGA}s,'' in \emph{2022 International Conference on Field-Programmable Technology (ICFPT)}, 2022, pp. 1--9.

\bibitem{resnet_og_paper}
\BIBentryALTinterwordspacing
K.~He, X.~Zhang, S.~Ren, and J.~Sun, ``Deep residual learning for image recognition,'' \emph{CoRR}, vol. abs/1512.03385, 2015. [Online]. Available: \url{http://arxiv.org/abs/1512.03385}
\BIBentrySTDinterwordspacing

\bibitem{jedec}
``{High Bandwidth Memory {DRAM} ({HBM2})},'' Joint Electron Device Engineering Council, Standard, Jan. 2016.

\bibitem{gpu_hbm}
M.~Zhu, Y.~Zhuo, C.~Wang, W.~Chen, and Y.~Xie, ``Performance evaluation and optimization of {HBM}-enabled {GPU} for data-intensive applications,'' in \emph{Design, Automation \& Test in Europe Conference \& Exhibition (DATE), 2017}, 2017, pp. 1245--1248.

\bibitem{shuhai}
H.~Huang, Z.~Wang, J.~Zhang, Z.~He, C.~Wu, J.~Xiao, and G.~Alonso, ``Shuhai: A tool for benchmarking high bandwidth memory on {FPGAs},'' \emph{IEEE Transactions on Computers}, vol.~71, no.~5, pp. 1133--1144, 2022.

\bibitem{vgg_og_paper}
K.~Simonyan and A.~Zisserman, ``Very deep convolutional networks for large-scale image recognition,'' 2015.

\bibitem{googlenet_og_paper}
C.~Szegedy, W.~Liu, Y.~Jia, P.~Sermanet, S.~Reed, D.~Anguelov, D.~Erhan, V.~Vanhoucke, and A.~Rabinovich, ``Going deeper with convolutions,'' 2014.

\bibitem{mv1_og_paper}
A.~G. Howard, M.~Zhu, B.~Chen, D.~Kalenichenko, W.~Wang, T.~Weyand, M.~Andreetto, and H.~Adam, ``{MobileNets}: Efficient convolutional neural networks for mobile vision applications,'' 2017.

\bibitem{mv2_og_paper}
M.~Sandler, A.~Howard, M.~Zhu, A.~Zhmoginov, and L.-C. Chen, ``{MobileNetV2}: Inverted residuals and linear bottlenecks,'' 2019.

\bibitem{mv3_og_paper}
A.~Howard, M.~Sandler, G.~Chu, L.-C. Chen, B.~Chen, M.~Tan, W.~Wang, Y.~Zhu, R.~Pang, V.~Vasudevan, Q.~V. Le, and H.~Adam, ``Searching for {MobileNetV3},'' 2019.

\bibitem{finn}
\BIBentryALTinterwordspacing
Y.~Umuroglu, N.~J. Fraser, G.~Gambardella, M.~Blott, P.~H.~W. Leong, M.~Jahre, and K.~A. Vissers, ``{FINN:} {A} framework for fast, scalable binarized neural network inference,'' \emph{CoRR}, vol. abs/1612.07119, 2016. [Online]. Available: \url{http://arxiv.org/abs/1612.07119}
\BIBentrySTDinterwordspacing

\bibitem{fpgaconvnet}
S.~I. Venieris and C.-S. Bouganis, ``Latency-driven design for fpga-based convolutional neural networks,'' in \emph{2017 27th International Conference on Field Programmable Logic and Applications (FPL)}, 2017, pp. 1--8.

\bibitem{satay}
A.~Montgomerie-Corcoran, P.~Toupas, Z.~Yu, and C.-S. Bouganis, ``{SATAY}: A streaming architecture toolflow for accelerating {YOLO} models on {FPGA} devices,'' in \emph{2023 International Conference on Field Programmable Technology (ICFPT)}, 2023, pp. 179--187.

\bibitem{traininghbm}
S.~K. Venkataramanaiah, H.-S. Suh, S.~Yin, E.~Nurvitadhi, A.~Dasu, Y.~Cao, and J.-S. Seo, ``{FPGA}-based low-batch training accelerator for modern {CNN}s featuring high bandwidth memory,'' in \emph{2020 IEEE/ACM International Conference On Computer Aided Design (ICCAD)}, 2020, pp. 1--8.

\bibitem{rwnn}
R.~Kuramochi and H.~Nakahara, ``An {FPGA}-based low-latency accelerator for randomly wired neural networks,'' in \emph{2020 30th International Conference on Field-Programmable Logic and Applications (FPL)}, 2020, pp. 298--303.

\bibitem{multicorehbm}
V.-C. Nguyen and Y.~Nakashima, ``Implementation of fully-pipelined {CNN} inference accelerator on {FPGA} and {HBM2} platform,'' \emph{IEICE Transactions on Information and Systems}, vol. 106, 2023.

\bibitem{stratix10}
\BIBentryALTinterwordspacing
D.~Lewis, G.~Chiu, J.~Chromczak, D.~Galloway, B.~Gamsa, V.~Manohararajah, I.~Milton, T.~Vanderhoek, and J.~Van~Dyken, ``The stratix™ 10 highly pipelined {FPGA} architecture,'' in \emph{Proceedings of the 2016 ACM/SIGDA International Symposium on Field-Programmable Gate Arrays}, ser. FPGA '16.\hskip 1em plus 0.5em minus 0.4em\relax New York, NY, USA: Association for Computing Machinery, 2016, p. 159–168. [Online]. Available: \url{https://doi.org/10.1145/2847263.2847267}
\BIBentrySTDinterwordspacing

\bibitem{carloni}
L.~P. Carloni, ``From latency-insensitive design to communication-based system-level design,'' \emph{Proceedings of the IEEE}, vol. 103, no.~11, pp. 2133--2151, 2015.

\bibitem{8bit_int_accuracy}
\BIBentryALTinterwordspacing
H.~Wu, P.~Judd, X.~Zhang, M.~Isaev, and P.~Micikevicius, ``Integer quantization for deep learning inference: Principles and empirical evaluation,'' \emph{CoRR}, vol. abs/2004.09602, 2020. [Online]. Available: \url{https://arxiv.org/abs/2004.09602}
\BIBentrySTDinterwordspacing

\bibitem{unzipfpga}
\BIBentryALTinterwordspacing
S.~I. Venieris, J.~Fernandez-Marques, and N.~D. Lane, ``Mitigating memory wall effects in {CNN} engines with on-the-fly weights generation,'' \emph{ACM Trans. Des. Autom. Electron. Syst.}, vol.~28, no.~6, oct 2023. [Online]. Available: \url{https://doi.org/10.1145/3611673}
\BIBentrySTDinterwordspacing

\bibitem{filmqnn}
\BIBentryALTinterwordspacing
M.~Sun, Z.~Li, A.~Lu, Y.~Li, S.-E. Chang, X.~Ma, X.~Lin, and Z.~Fang, ``{FILM-QNN}: Efficient {FPGA} acceleration of deep neural networks with intra-layer, mixed-precision quantization,'' in \emph{Proceedings of the 2022 ACM/SIGDA International Symposium on Field-Programmable Gate Arrays}, ser. FPGA '22.\hskip 1em plus 0.5em minus 0.4em\relax New York, NY, USA: Association for Computing Machinery, 2022, p. 134–145. [Online]. Available: \url{https://doi.org/10.1145/3490422.3502364}
\BIBentrySTDinterwordspacing

\bibitem{bestresnet50}
S.~Liu, H.~Fan, M.~Ferianc, X.~Niu, H.~Shi, and W.~Luk, ``Toward full-stack acceleration of deep convolutional neural networks on {FPGAs},'' \emph{IEEE Transactions on Neural Networks and Learning Systems}, vol.~33, no.~8, pp. 3974--3987, 2022.

\bibitem{dnnvm}
\BIBentryALTinterwordspacing
Y.~Xing, S.~Liang, L.~Sui, Z.~Zhang, J.~Qiu, X.~Jia, X.~Liu, Y.~Wang, Y.~Shan, and Y.~Wang, ``{DNNVM}: End-to-end compiler leveraging operation fusion on {FPGA}-based {CNN} accelerators,'' in \emph{Proceedings of the 2019 ACM/SIGDA International Symposium on Field-Programmable Gate Arrays}, ser. FPGA '19.\hskip 1em plus 0.5em minus 0.4em\relax New York, NY, USA: Association for Computing Machinery, 2019, p. 187–188. [Online]. Available: \url{https://doi.org/10.1145/3289602.3293972}
\BIBentrySTDinterwordspacing

\bibitem{ftdl}
R.~Shi, Y.~Ding, X.~Wei, H.~Li, H.~Liu, H.~K.-H. So, and C.~Ding, ``{FTDL}: A tailored {FPGA}-overlay for deep learning with high scalability,'' in \emph{2020 57th ACM/IEEE Design Automation Conference (DAC)}, 2020, pp. 1--6.

\bibitem{bnn_batchone}
\BIBentryALTinterwordspacing
``Resnet-50 pynq github,'' (Date last accessed 14-March-2024). [Online]. Available: \url{https://github.com/Xilinx/ResNet50-PYNQ/blob/master/host/README.md}
\BIBentrySTDinterwordspacing

\bibitem{fpgaconvnet_vgg}
S.~I. Venieris and C.-S. Bouganis, ``fpgaconvnet: Mapping regular and irregular convolutional neural networks on {FPGA}s,'' \emph{IEEE Transactions on Neural Networks and Learning Systems}, vol.~30, no.~2, pp. 326--342, 2019.

\bibitem{best_vgg16}
Y.~Ma, Y.~Cao, S.~Vrudhula, and J.-S. Seo, ``Automatic compilation of diverse {CNN}s onto high-performance {FPGA} accelerators,'' \emph{IEEE Transactions on Computer-Aided Design of Integrated Circuits and Systems}, vol.~39, no.~2, pp. 424--437, 2020.

\bibitem{hbm_interconnect}
\BIBentryALTinterwordspacing
Y.-k. Choi, Y.~Chi, W.~Qiao, N.~Samardzic, and J.~Cong, ``{HBM Connect}: High-performance {HLS} interconnect for {FPGA} {HBM},'' in \emph{The 2021 ACM/SIGDA International Symposium on Field-Programmable Gate Arrays}, ser. FPGA '21.\hskip 1em plus 0.5em minus 0.4em\relax New York, NY, USA: Association for Computing Machinery, 2021, p. 116–126. [Online]. Available: \url{https://doi.org/10.1145/3431920.3439301}
\BIBentrySTDinterwordspacing

\end{thebibliography}

\end{document}